\documentclass[11pt]{article}

\usepackage{acl}

\usepackage{multirow}
\usepackage[table]{xcolor}
\usepackage{array}
\usepackage{graphicx}
\usepackage{makecell}
\usepackage{graphicx}
\usepackage{subcaption}
\usepackage{amsfonts}
\usepackage{amssymb}

\usepackage{times}
\usepackage{latexsym}
\usepackage{multirow}
\usepackage[T1]{fontenc}

\usepackage[utf8]{inputenc}

\usepackage{microtype}

\usepackage{inconsolata}

\usepackage{graphicx}

\title{OrgAgent: Organize Your Multi-Agent System like a Company}

\author{
  Yiru Wang\textsuperscript{1} 
  Xinyue Shen\textsuperscript{3} 
  Yaohui Han\textsuperscript{1} 
  Michael Backes\textsuperscript{3}
  Pin-Yu Chen\textsuperscript{2}
  Tsung-Yi Ho\textsuperscript{1} \\[0.8em]
  \textsuperscript{1}The Chinese University of Hong Kong, 
  \textsuperscript{2}IBM Research \\
  \textsuperscript{3}CISPA Helmholtz Center for Information Security \\
}

\begin{document}
\maketitle

\begin{abstract}
While large language model-based multi-agent systems have shown strong potential for complex reasoning, how to effectively organize multiple agents remains an open question.
In this paper, we introduce \textbf{OrgAgent}, a company-style hierarchical multi-agent framework that separates collaboration into governance, execution, and compliance layers. 
OrgAgent decomposes multi-agent reasoning into three layers: a governance layer for planning and resource allocation, an execution layer for task solving and review, and a compliance layer for final answer control.
By evaluating the framework across reasoning tasks, LLMs, execution modes, and execution policies, we find that multi-agent systems organized in a company-style hierarchy generally outperform other organizational structures.
Besides, hierarchical coordination also reduces token consumption relative to flat collaboration in most settings. 
For example, for GPT-OSS-120B, the hierarchical setting improves performance over flat multi-agent system by 102.73\% while reducing token usage by 74.52\% on SQuAD 2.0.
Further analysis shows that hierarchy helps most when tasks benefit from stable skill assignment, controlled information flow, and layered verification.
Overall, our findings highlight organizational structure as an important factor in multi-agent reasoning, shaping not only effectiveness and cost, but also coordination behavior.
\end{abstract}

\section{Introduction}

\begin{figure}[t]
    \centering
    \includegraphics[width=\linewidth]{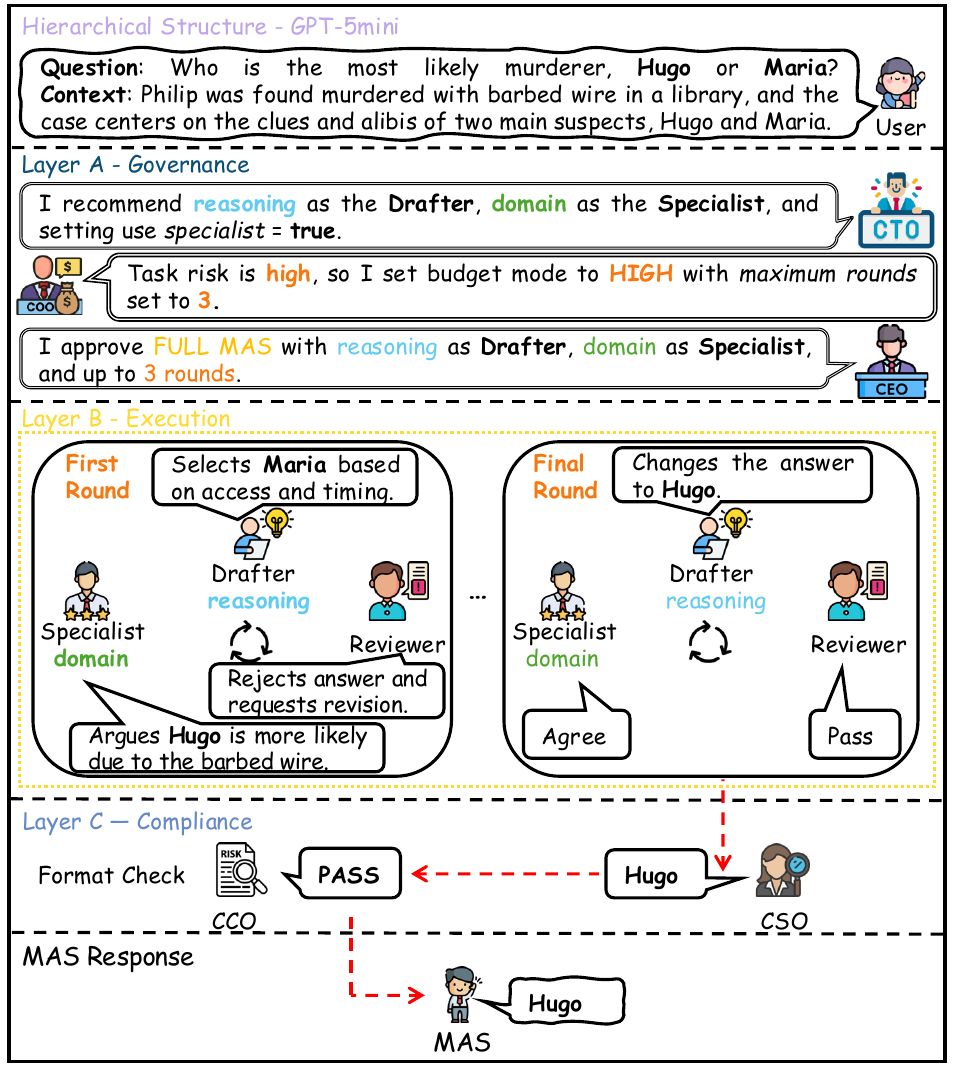}
    \caption{
    Illustration of our company-style hierarchical MAS framework \textbf{OrgAgent}. 
    Layer A performs governance-level planning, including skill assignment and execution control; Layer B carries out task solving through collaborative drafting and feedback; Layer C finalizes the output through answer consolidation and compliance checking.}
    \label{fig:case}
\end{figure}

Large language models (LLMs) have evolved from single-turn assistants into increasingly autonomous agents capable of planning, tool use, and collaboration. These advances have driven the development of \textbf{LLM-based Multi-Agent Systems (MAS)}, which are increasingly studied in complex settings such as problem solving, software engineering, and simulation~\cite{guo2024large,li2024survey,he2025llm}. 
Existing research has developed along two directions. 
One line studies \textit{interaction mechanisms} among agents, focusing on how agents communicate and coordinate through role-playing, discussion, debate, voting, or consensus, as exemplified by CAMEL~\cite{li2023camel}. 
The other line focuses on \textit{higher-level organization}, emphasizing role assignment, workflow design, and system-level coordination, as represented by frameworks such as AutoGen~\cite{wu2024autogen} and role-specialized collaborative systems including MetaGPT~\cite{hong2023metagpt}, ChatDev~\cite{qian2024chatdev}, and Paperclip~\cite{paperclip2026}.

One natural way to organize MAS is through organizational structure~\cite{pugh1971organization,mintzberg1979structuring,daft2007organization}. In organizational theory, organizational structure determines how tasks, coordination, supervision, and decision authority are distributed, thereby shaping organizational behavior~\cite{burton2012organisational}.
Common forms include flat structures~\cite{ghiselli1972leadership} with fewer managerial layers and hierarchical structures~\cite{child2019hierarchy} with more complex management.
Among these, the company-style hierarchy has been refined over decades, developing well-established mechanisms for goal alignment, role division, resource allocation, and outcome verification~\cite{mintzberg1979structuring,burton2012organisational}. 
This makes company-style hierarchy a natural basis for organizing MAS, as it explicitly defines who plans, who executes, who reviews, and how decisions are controlled. 

In this work, as shown in ~\autoref{fig:case}, we instantiate organizational structure as a company style hierarchy, one of its most common real-world representations, to study how structured governance affects multi-agent reasoning.
OrgAgent decomposes the reasoning process into three layers: 1) a \textit{governance} layer for planning, routing, and resource allocation; 2) an \textit{execution} layer for answer generation, critique, and revision, whose interaction process is further controlled through different execution modes and execution policies; 3) and a \textit{compliance} layer for final answer validation and output control.
We then evaluate the framework on three reasoning benchmarks, MuSR, MuSiQue, and SQuAD 2.0, using three language models and multiple execution modes and policies. 
Results show that MAS organized in the company-style hierarchy generally outperforms both flat and single-agent baselines, with especially clear gains on MuSiQue and SQuAD 2.0.

Our main contributions are as follows:
\begin{itemize}
    \item We propose \textbf{OrgAgent}, a company-style hierarchical MAS framework that separates governance, execution, and compliance, supported by a skill-based worker pool and various execution modes and policies.
    \item We present the first systematic empirical study of flat and hierarchical MAS on general reasoning tasks, treating organizational structure itself as the central variable of analysis.
    \item We show that company-style hierarchy improves both effectiveness and efficiency in most settings, often achieving higher task performance while reducing token cost, with gains of up to +102.73\% in F1-score and 74.52\% fewer tokens on SQuAD 2.0.
    \item We provide a fine-grained analysis of coordination behavior, showing when hierarchical organization is effective and when it may be limited by additional overhead or coordination constraints.
\end{itemize}

\section{Related Work}

\paragraph{Organizational Structure.}

Organizational structure is a core concept in organization theory and organizational design~\cite{joseph2025organization,mintzberg1979structuring,burton2012organisational,daft2007organization}. Classic and contemporary work studies how different structural forms distribute authority, coordination, and specialization, including functional, hierarchical, and matrix arrangements~\cite{duncan1979right,galbraith1971matrix,kates2010designing,anand2007right}. Recent reviews further identify configuration, control, channelization, and coordination as major perspectives in organization design research~\cite{joseph2025organization}.
Empirical studies also show that structure matters in practice, for example, in healthcare quality~\cite{hearld2008hospital} and manufacturing innovation and operational performance~\cite{iranmanesh2021impacts}. 
However, this literature primarily concerns human organizations and does not explain how structure should be instantiated for within-task governance in LLM-based multi-agent systems.

\begin{figure*}[t]
    \centering
    \includegraphics[width=\textwidth]{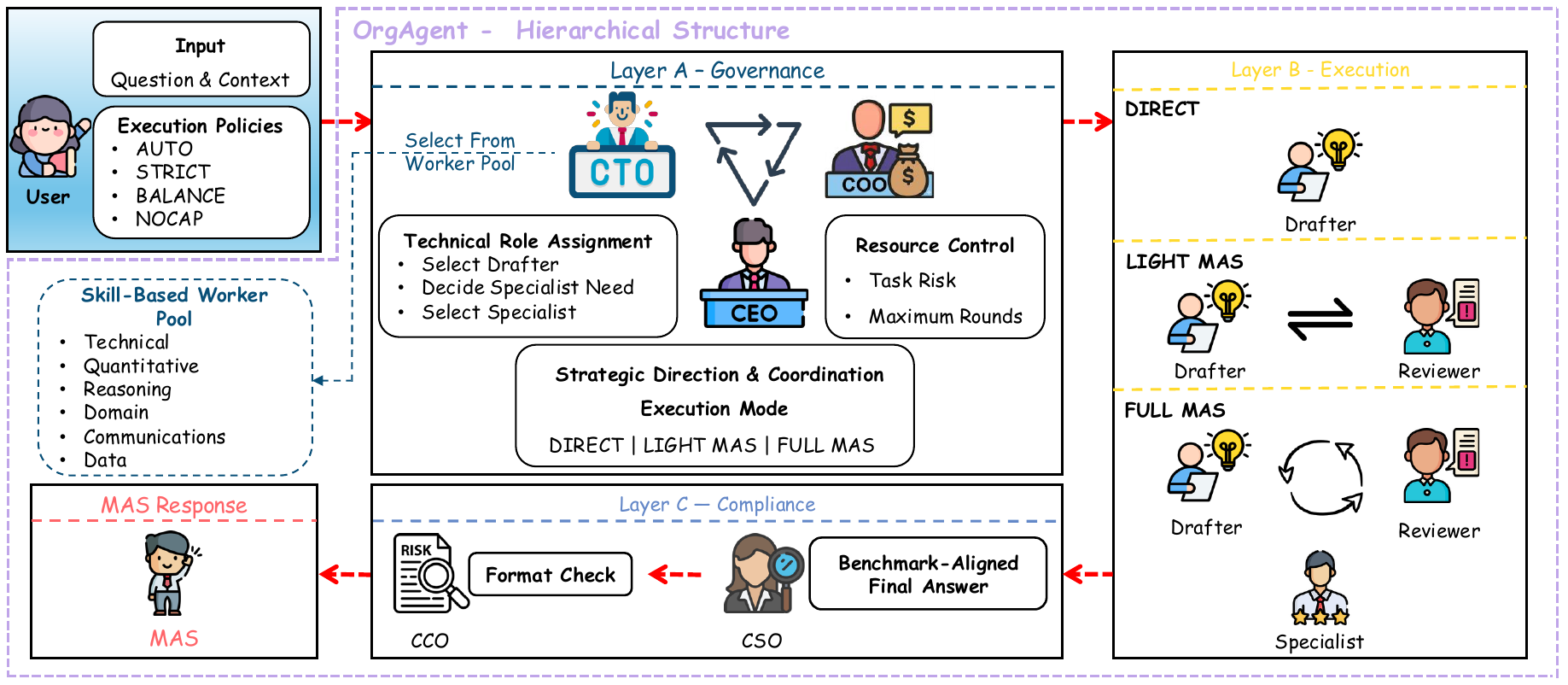}
    \caption{
    Overview of \textbf{OrgAgent}, a company-style hierarchical MAS framework.
    }
    \label{fig:flow}
\end{figure*}
\paragraph{LLM-Based MAS.}

Prior work on LLM-based MAS mainly falls into two lines~\cite{guo2024large}. One line studies local interaction and decision mechanisms. CAMEL uses role playing for autonomous cooperation~\cite{li2023camel}. Multi-agent Debate uses iterative critique to improve reasoning~\cite{du2023improving}. Voting or Consensus shows that decision protocol, agent count, and discussion rounds can substantially affect performance~\cite{kaesberg2025voting}. AgentVerse studies collaborative group composition and emergent behaviors~\cite{chen2023agentverse}. The other line studies higher-level orchestration. AutoGen provides a general infrastructure for multi-agent conversation~\cite{wu2024autogen}. MetaGPT organizes agents through role specialization and standardized procedures~\cite{hong2023metagpt}. ChatDev structures software development through specialized agents and chat chains~\cite{qian2024chatdev}. \textit{From Bits to Boardrooms} proposes a hierarchical framework linking operational analysis with strategic decision making~\cite{wang2025bits}. However, it remains unclear how organizational structure itself should govern within task coordination in MAS, which our work addresses.

\section{OrgAgent}

In this section, we introduce \textbf{OrgAgent}, as shown in~\autoref{fig:flow}, a multi-agent framework that organizes collaboration through three layers of governance, execution, and compliance. We describe its architecture, agent roles, and execution policies, which together define how coordination is structured within a single task.

\subsection{Organizational Structures in Management}

Organizational structure refers to how authority, roles, and coordination are arranged within an organization. In management research, different structures shape how decisions are made and how work is controlled~\cite{mintzberg1979structuring}. In this paper, we focus on two common forms in management: \textbf{flat organization} and a \textbf{hierarchical organization}, because they represent two contrasting ways of organizing collective work.

\paragraph{Flat Organization.}
A flat organization refers to a flatter structure with reduced vertical differentiation. It is usually characterized by shorter communication paths, less layered supervision, and greater autonomy. Its main advantage is flexibility~\cite{reitzig2022get,lee2022myth}, while its main limitation is weaker control and less clear coordination in complex tasks.

\paragraph{Hierarchical Organization.}

A hierarchical organization refers to a structure with multiple levels of authority and clearer reporting relationships. It is characterized by stronger supervision, clearer role differentiation, and more structured coordination. Its main advantage is control and accountability~\cite{halevy2011functional}, while its main limitation is slower communication and lower flexibility.
\subsection{Agents}
We define the agents used in our system. The same set of agents can be reused under different organizational settings, while their interaction patterns may vary across frameworks.

\paragraph{Chief Executive Officer (CEO).}
The CEO~\cite{vancil1987passing} focuses on overall strategic direction and high-level coordination. Its role is to keep the problem-solving process aligned with the overall objective of the task.

\paragraph{Chief Technology Officer (CTO).}
The CTO~\cite{medcof2007cto} focuses on technical soundness and solution design. Its role is to examine the technical direction of the solution and ensure that the problem-solving process remains technically appropriate.

\paragraph{Chief Operating Officer (COO).}
The COO~\cite{bennett2020riding} focuses on operational resources and execution efficiency. Its role is to consider resource usage, execution constraints, and the overall efficiency of the process.

\paragraph{Drafter.}
The Drafter is the primary writer in the problem-solving process. Its role is to produce the main candidate answer and revise it when necessary.

\paragraph{Reviewer.}
The Reviewer focuses on answer quality and error detection. Its role is to examine the current draft, identify potential weaknesses or inconsistencies, and determine whether revision is needed.

\paragraph{Specialist.}
The Specialist focuses on targeted support for difficult or error-prone parts of the task. Its role is to provide additional expertise or refinement when the main draft requires further support.

\paragraph{Chief Solutions Officer (CSO).}
The CSO~\cite{wiki_cso} is responsible for producing the final answer under benchmark-specific constraints. Its role is to ensure that the final response matches the required answer format and task requirements of the target benchmark.

\paragraph{Chief Compliance Officer (CCO).}
The CCO~\cite{bcsc_cco} is responsible for checking whether the final output satisfies predefined structural requirements. Its role is to verify compliance with the required schema or output format, but it does not perform task reasoning itself.

\subsection{Skill-Based Worker Pool}

Our framework maintains a pool of six skill-based workers, which can serve as either the Drafter or the Specialist during execution.

\begin{itemize}
    \item \textbf{Technical}: Focuses on implementation details, procedural constraints, and structured problem solving.
    \item \textbf{Quantitative}: Focuses on numerical, symbolic, and stepwise reasoning.
    \item \textbf{Reasoning}: Focuses on logical consistency, multi-step inference, and chain coherence;
    \item \textbf{Domain}: Focuses on domain-specific interpretation and contextual understanding.
    \item \textbf{Communications}: Focuses on clarity, concise final phrasing, and answer presentation.
    \item \textbf{Data}: Focuses on evidence extraction, pattern recognition, and information organization.
\end{itemize}

These skill profiles are not tied to fixed benchmark types. Instead, they provide reusable capability orientations that can be instantiated under different execution roles depending on task needs.

\subsection{Flat Framework}
We implement a flat framework in which all participating agents operate at the same organizational level. Specifically, the \textbf{CEO}, \textbf{CTO}, \textbf{COO}, \textbf{Drafter}, \textbf{Reviewer}, \textbf{Specialist}, and \textbf{CSO} interact as peer agents without an explicit layered chain of command, and all of them work on the basis of shared task information and shared interaction context. Although these agents have different functional responsibilities, coordination, problem solving, answer checking, and final response generation are carried out within a single-level collaborative process. The \textbf{CCO} is not treated as a deliberative peer, but is used only for final structural compliance checking.

\subsection{Hierarchical Framework}
Our hierarchical framework \textbf{OrgAgent} organizes agents into three vertical layers, namely Layer A, Layer B, and Layer C. This design separates high-level coordination, task execution, and final output control into different stages, so that the problem-solving process follows a structured top-down workflow rather than a single-level interaction process.

\paragraph{Layer A (Governance Layer).}
Layer A is responsible for high-level coordination and planning. It includes the \textbf{CEO}, \textbf{CTO}, and \textbf{COO}, which respectively focus on strategic direction, technical direction, and operational resources. Based on the task input, this layer determines the execution configuration for the downstream process.

\paragraph{Layer B (Execution Layer).}
Layer B is responsible for task solving under the configuration determined by Layer A. It includes the \textbf{Drafter}, \textbf{Reviewer}, and, when needed, the \textbf{Specialist}. In this layer, the Drafter produces the candidate answer, the Reviewer checks its quality, and the Specialist provides targeted support for difficult or error-prone parts of the task.

\paragraph{Layer C (Compliance Layer).}
Layer C is responsible for final output generation and structural verification. It includes the \textbf{CSO}, which produces the final answer under benchmark-specific constraints, and the \textbf{CCO}, which checks whether the output satisfies the required structural format. In this way, the final response is both benchmark-aligned and structurally compliant.

\subsection{Execution Modes}
Our framework supports three execution modes: \textbf{DIRECT}, \textbf{LIGHT MAS}, and \textbf{FULL MAS}.
These modes differ in how the execution layer is organized and therefore provide different trade-offs between efficiency, verification strength, and coordination cost.

\paragraph{DIRECT.}
In DIRECT configuration, the execution layer relies on the \textbf{Drafter} to produce the candidate answer directly, without invoking additional review or specialist support. This mode minimizes execution overhead and is suitable for relatively simple tasks or resource-constrained settings.

\paragraph{LIGHT MAS.}
The LIGHT MAS configuration activates the \textbf{Drafter} and the \textbf{Reviewer}. In this setting, the Drafter first produces a candidate answer, and the Reviewer then checks its quality and determines whether revision is needed. Compared with DIRECT, this mode introduces an additional verification step while keeping the coordination cost relatively low.

\paragraph{FULL MAS.}
The FULL MAS configuration activates the \textbf{Drafter}, \textbf{Reviewer}, and \textbf{Specialist}. In addition to answer generation and review, this mode allows targeted expert support for difficult or error-prone parts of the task. As a result, it provides the strongest execution support, but also incurs the highest coordination and computation cost.

\subsection{Execution Policies}
Our framework supports four execution policies, namely \textbf{STRICT}, \textbf{BALANCE}, \textbf{NOCAP}, and \textbf{AUTO}.
These policies control how aggressively the framework constrains resource usage and collaboration during execution.

\paragraph{STRICT.}
The strict policy emphasizes conservative execution by imposing tighter resource and interaction constraints.

\paragraph{BALANCE.}
The balance policy provides a middle ground between efficiency and execution support.

\paragraph{NOCAP.}
The no-cap policy minimizes explicit execution constraints and allows more flexible resource usage when needed.

\paragraph{AUTO.}
The auto policy adaptively selects an execution configuration according to task characteristics.

\section{Experimental Setup}
In this section, we describe the experimental setup used to evaluate \textbf{OrgAgent}.

\subsection{Models}

We evaluate our framework with three backbone LLMs: \textbf{GPT-OSS-120B}~\cite{agarwal2025gpt}, \textbf{GPT-5 mini}~\cite{openai_gpt5_mini}, and \textbf{Llama 3.1 8B}~\cite{ollama_llama31}. These models represent different levels of capability, allowing us to examine whether the impact of organizational structure is universal across different models.

\subsection{Benchmarks}

We evaluate the framework on \textbf{MuSR}, \textbf{MuSiQue}, and \textbf{SQuAD 2.0}, which cover different forms of reasoning difficulty. Additional benchmark details are provided in Appendix~\ref{sec:benchmark_details}.

\paragraph{MuSR~\cite{sprague2023musr}} is a benchmark for multistep soft reasoning over long narratives, and we report \textbf{accuracy} as the evaluation metric.

\paragraph{MuSiQue~\cite{trivedi2022musique}} is a benchmark for compositional multi-hop question answering, and we report standard \textbf{F1} scores.

\paragraph{SQuAD 2.0~\cite{li2018question}} is a reading comprehension benchmark containing both answerable and unanswerable questions, and we report standard \textbf{F1} scores.

\begin{table*}[t]
\centering

\resizebox{\textwidth}{!}{%
\begin{tabular}{l|c|c|c|c|c|c|c|c}

\hline
\multirow{2}{*}{\textbf{Model}}
& \multicolumn{2}{c|}{\textbf{Baseline}}
& \multicolumn{2}{c|}{\textbf{Flat}}
& \multicolumn{2}{c|}{\textbf{Hierarchical (AUTO)}}
& \multirow{2}{*}{\shortstack{\textbf{$\Delta$ Improvement}\\\textbf{(\%)}}}
& \multirow{2}{*}{\shortstack{\textbf{$\Delta$ Token}\\\textbf{Reduction (\%)}}} \\
\cline{2-7}
& \textbf{Score} & \textbf{Avg token}
& \textbf{Score} & \textbf{Avg token}
& \textbf{Score} & \textbf{Avg token}
& & \\
\hline

\multicolumn{9}{>{\columncolor{gray!20}}c}{\textbf{MuSiQue (F1-score)}} \\
\hline
GPT-5mini    & $51.28 \pm 4.2$             & 2,778  & $50.31 \pm 2.50$          & 28,479 & $\mathbf{68.98 \pm 1.70}$  & 11,408 & +37.11\%  & 59.94\% \\
GPT-OSS-120B & $37.98 \pm 2.58$            & 2,687  & $48.40 \pm 1.55$          & 25,209 & $\mathbf{57.58 \pm 1.98}$  & 12,046 & +18.97\%  & 52.22\% \\
LLaMA-3.1-8B & $11.52 \pm 3.09$            & 2,370  & $14.55 \pm 0.09$          & 51,425 & $\mathbf{32.59 \pm 14.65}$ & 12,322 & +123.99\% & 76.04\% \\
\hline

\multicolumn{9}{>{\columncolor{gray!20}}c}{\textbf{MuSR (Accuracy)}} \\
\hline
GPT-5mini    & $29.00 \pm 1.41$            & 1,603  & $62.45 \pm 5.80$          & 13,419 & $\mathbf{64.83 \pm 2.87}$  & 7,195  & +3.81\%   & 46.38\% \\
GPT-OSS-120B & $50.65 \pm 2.94$            & 1,328  & $\mathbf{69.00     \pm 1.54}$          & 12,700 & $59.50 \pm 1.08$           & 5,994  & -13.77\%  & 52.80\% \\
LLaMA-3.1-8B & $10.33 \pm 2.49$            & 1,061  & $\mathbf{37.41 \pm 1.09}$ & 25,600 & $34.00 \pm 0.71$           & 5,912  & -9.12\%   & 76.91\% \\
\hline

\multicolumn{9}{>{\columncolor{gray!20}}c}{\textbf{SQuAD 2.0 (F1-score)}} \\
\hline
GPT-5mini    & $31.34 \pm 0.95$            & 458    & $28.77 \pm 3.07$          & 15,683 & $\mathbf{63.43 \pm 1.51}$  & 3,245  & +120.47\% & 79.31\% \\
GPT-OSS-120B & $26.61 \pm 1.60$            & 425    & $31.12 \pm 0.03$          & 13,021 & $\mathbf{63.09 \pm 1.52}$  & 3,318  & +102.73\% & 74.52\% \\
LLaMA-3.1-8B & $24.92 \pm 2.62$            & 240    & $28.17 \pm 2.94$          & 22,806 & $\mathbf{44.78 \pm 3.03}$  & 5,188  & +58.96\%  & 77.25\% \\
\hline
\end{tabular}%
}
\caption{Performance and token cost comparison across baseline, flat, and hierarchical organizations.}
\label{tab:main_results}
\end{table*}

\subsection{Evaluation Metrics}

We evaluate each system from three perspectives: \textbf{task performance}, \textbf{token efficiency}, and \textbf{coordination behavior}.

\paragraph{Task Performance.}
For \textbf{MuSR}, we report \textbf{Accuracy}. For \textbf{MuSiQue} and \textbf{SQuAD 2.0}, we report the standard \textbf{F1-score}. Let $N$ denote the total number of evaluation examples and $K$ the number of repeated runs for each setting. Since the framework is stochastic, each setting is run multiple times, and we report the mean and standard deviation across runs. Detailed definitions of the benchmark-specific metrics are provided in Appendix~\ref{app:benchmark_metrics}.

\paragraph{Token Efficiency.}
To measure coordination cost, we report the average token usage per example:
\begin{equation}
\mathrm{AvgToken}=\frac{1}{N}\sum_{i=1}^{N} t_i,
\end{equation}
where $i$ indexes an evaluation example, and $t_i$ is the total number of tokens consumed for example $i$, including all agent interactions and final answer generation.

To compare hierarchical and flat coordination, we further compute relative score improvement and token reduction:
\begin{equation}
Improvement(\%)=
\frac{S_{\mathrm{hier}}-S_{\mathrm{flat}}}{S_{\mathrm{flat}}}\times 100,
\end{equation}
\begin{equation}
Token Reduction(\%)=
\frac{T_{\mathrm{flat}}-T_{\mathrm{hier}}}{T_{\mathrm{flat}}}\times 100,
\end{equation}
where $S_{\mathrm{hier}}$ and $S_{\mathrm{flat}}$ denote the final task performance scores of the hierarchical and flat settings, respectively, and $T_{\mathrm{hier}}$ and $T_{\mathrm{flat}}$ denote their average token usage.

\paragraph{Coordination Behavior.}
We analyze coordination behavior through the distribution of selected skill types for the \textbf{Drafter} and \textbf{Specialist} roles. For the unanswerable subset of \textbf{SQuAD 2.0}, we also report the \textbf{abstention rate}, defined as the proportion of unanswerable examples for which the system outputs a normalized no-answer response. Detailed definitions are provided in Appendix~\ref{app:benchmark_metrics}.

\section{Results}
In this section, we conduct extensive experiments to evaluate the effectiveness and coordination behavior of our organizationally structured multi-agent framework. 
We aim to address the following research questions: 
1) In general reasoning tasks, can hierarchical organization outperform flat MAS and single-agent baselines? 
2) How do different organizational structures, execution modes, and execution policies trade off task accuracy against token cost? 
3) What coordination patterns emerge under different organizational settings? 

\subsection{Performance of Different Structures}

We present the quantitative comparison of performance and token cost across baseline, flat, and hierarchical organizations in~\autoref{tab:main_results}. The results indicate that hierarchical organization generally outperforms both flat MAS and single-agent baselines. This advantage is especially clear on MuSiQue and SQuAD 2.0, where the hierarchical setting achieves the best performance for all three models. 
On MuSiQue, hierarchical organization improves F1-score by 37.11\% for GPT-5mini and 123.99\% for LLaMA-3.1-8B, while also bringing an 18.97\% gain for GPT-OSS-120B. On SQuAD 2.0, the gains are even larger, reaching 120.47\%, 102.73\%, and 58.96\% over flat MAS for GPT-5mini, GPT-OSS-120B, and LLaMA-3.1-8B, respectively.

\begin{figure*}[t]
    \centering
    \includegraphics[width=\textwidth]{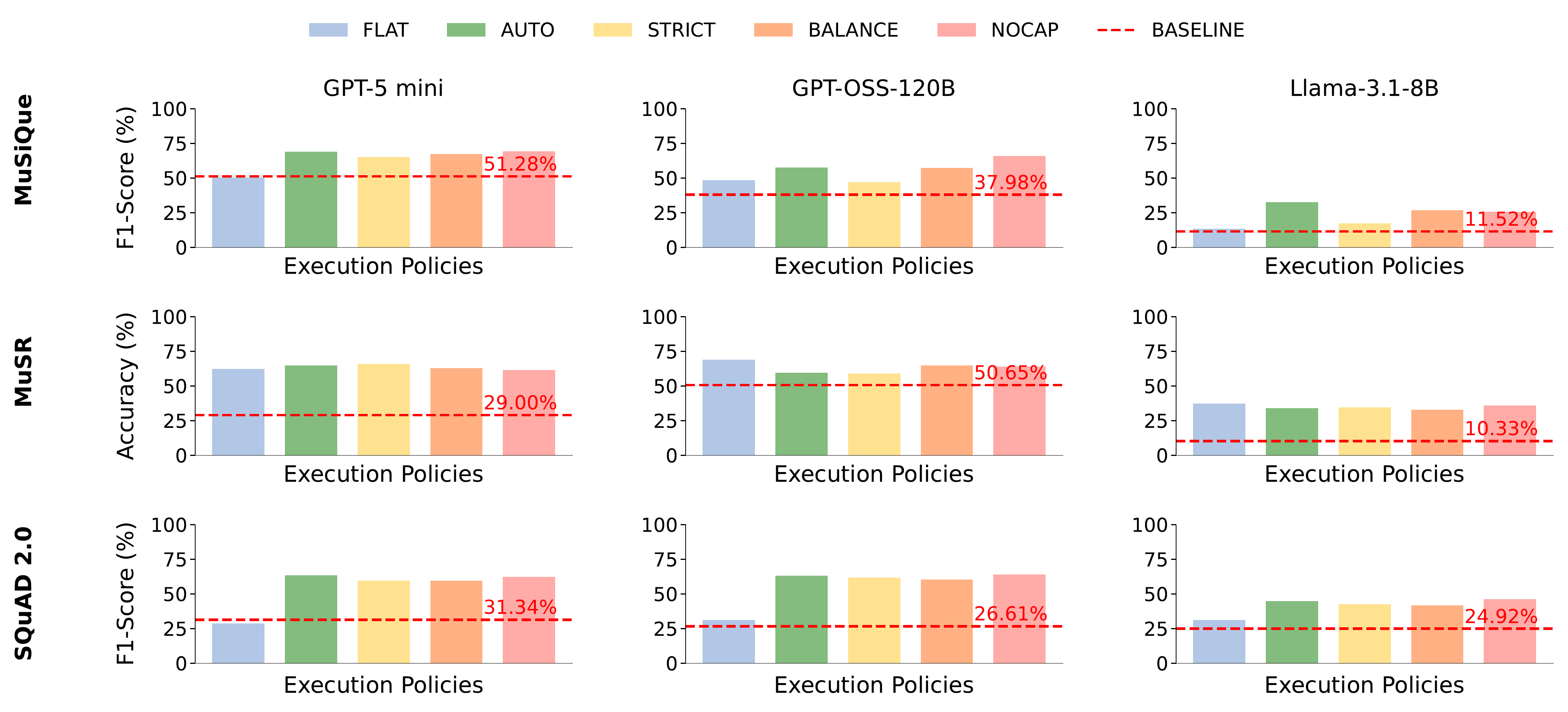}
    \caption{Performance comparison of different execution policies across three benchmarks. Rows correspond to MuSiQue, MuSR, and SQuAD 2.0, while columns correspond to GPT-5 mini, GPT-OSS-120B, and Llama-3.1-8B. Bars denote the performance under FLAT, AUTO, STRICT, BALANCE, and NOCAP policies, and the red dashed line indicates the single-agent baseline.}
    \label{fig:three_bench_3x3}
\end{figure*}

The results on MuSR are more mixed.
Hierarchical organization slightly outperforms flat MAS for GPT-5mini, but remains below the flat setting for GPT-OSS-120B and LLaMA-3.1-8B.
This suggests that hierarchical coordination is not uniformly dominant across all reasoning tasks. 
At the same time, these two cases also show that when hierarchical coordination fails to translate additional structure into better answer quality, it may still require substantial coordination cost in tokens.

More importantly, hierarchical organization uses substantially fewer tokens than flat MAS in every setting. Compared with the flat organization, token usage decreases consistently across all three benchmarks and all three models, with reductions ranging from 46.38\% to 79.31\%. This reduction is not marginal, but large and systematic: the hierarchical framework never increases token cost relative to flat MAS, and instead consistently cuts interaction overhead by nearly half or more. Overall, these results show that introducing explicit layers of governance, execution, and compliance can improve coordination quality while substantially reducing the communication cost of multi-agent reasoning.

\subsection{Accuracy and Token Cost Trade off}

We compare different execution policies within the hierarchical framework in terms of task performance in~\autoref{fig:three_bench_3x3} and average token consumption in~\autoref{tab:avg_token_exec_only}. We first examine how performance varies across policies, as shown in~\autoref{fig:three_bench_3x3}, different execution policies lead to distinct performance patterns across benchmarks and models. On MuSiQue, AUTO, BALANCE, and NOCAP generally achieve stronger F1-scores than STRICT, indicating that allowing more flexible coordination is beneficial for this benchmark. In particular, for GPT-5 mini and GPT-OSS-120B, the best results are obtained under AUTO or BALANCE, while for Llama-3.1-8B, the execution policies all remain clearly above the baseline and flat setting. On MuSR, the differences among policies are smaller for GPT-5 mini, but become more visible for GPT-OSS-120B and Llama-3.1-8B. In these cases, no single policy dominates across all models, suggesting that policy effectiveness is more model-dependent on this benchmark. On SQuAD 2.0, the performance gap among execution policies is relatively small, and all of them remain substantially stronger than the baseline and flat setting. This suggests that for SQuAD 2.0, the main benefit comes from adopting structured hierarchical coordination itself, while the choice of policy mainly affects efficiency rather than final accuracy.

\begin{table}[t]
\centering
\small
\renewcommand{\arraystretch}{1.15}
\setlength{\tabcolsep}{5pt}
\resizebox{\columnwidth}{!}{%
\begin{tabular}{l|c|c|c|c|c}
\hline
\multicolumn{6}{c}{\textbf{Avg Token}} \\
\hline
Model & AUTO & STRICT & BALANCE & NOCAP & FLAT \\
\hline
\rowcolor{gray!20}
\multicolumn{6}{c}{\textbf{MuSiQue}} \\
\hline
GPT-5mini    & 11,545 & 3,795 & 12,711 & 21,766 & 28,479 \\
GPT-OSS-120B & 12,046 & 3,633 & 12,428 & 14,539 & 25,209 \\
LLaMA-3.1-8B & 12,322 & 3,282 & 12,399 & 37,306 & 48,198 \\
\hline
\rowcolor{gray!20}
\multicolumn{6}{c}{\textbf{MuSR}} \\
\hline
GPT-5mini    & 7,195 & 3,275 & 7,128 & 9,294 & 13,419 \\
GPT-OSS-120B & 5,994 & 3,128 & 6,953 & 7,063 & 12,700 \\
LLaMA-3.1-8B & 5,912 & 2,176 & 7,691 & 16,639 & 25,600 \\
\hline
\rowcolor{gray!20}
\multicolumn{6}{c}{\textbf{SQuAD 2.0}} \\
\hline
GPT-5mini    & 3,245 & 1,554 & 3,215 & 3,513 & 15,683 \\
GPT-OSS-120B & 3,318 & 1,539 & 3,156 & 3,574 & 13,021 \\
LLaMA-3.1-8B & 5,188 & 1,148 & 3,674 & 11,385 & 22,806 \\
\hline
\end{tabular}%
}
\caption{Average token consumption under different execution policies across benchmarks and models.}
\label{tab:avg_token_exec_only}
\end{table} 

\begin{figure*}[t]
    \centering
    \includegraphics[width=\linewidth]{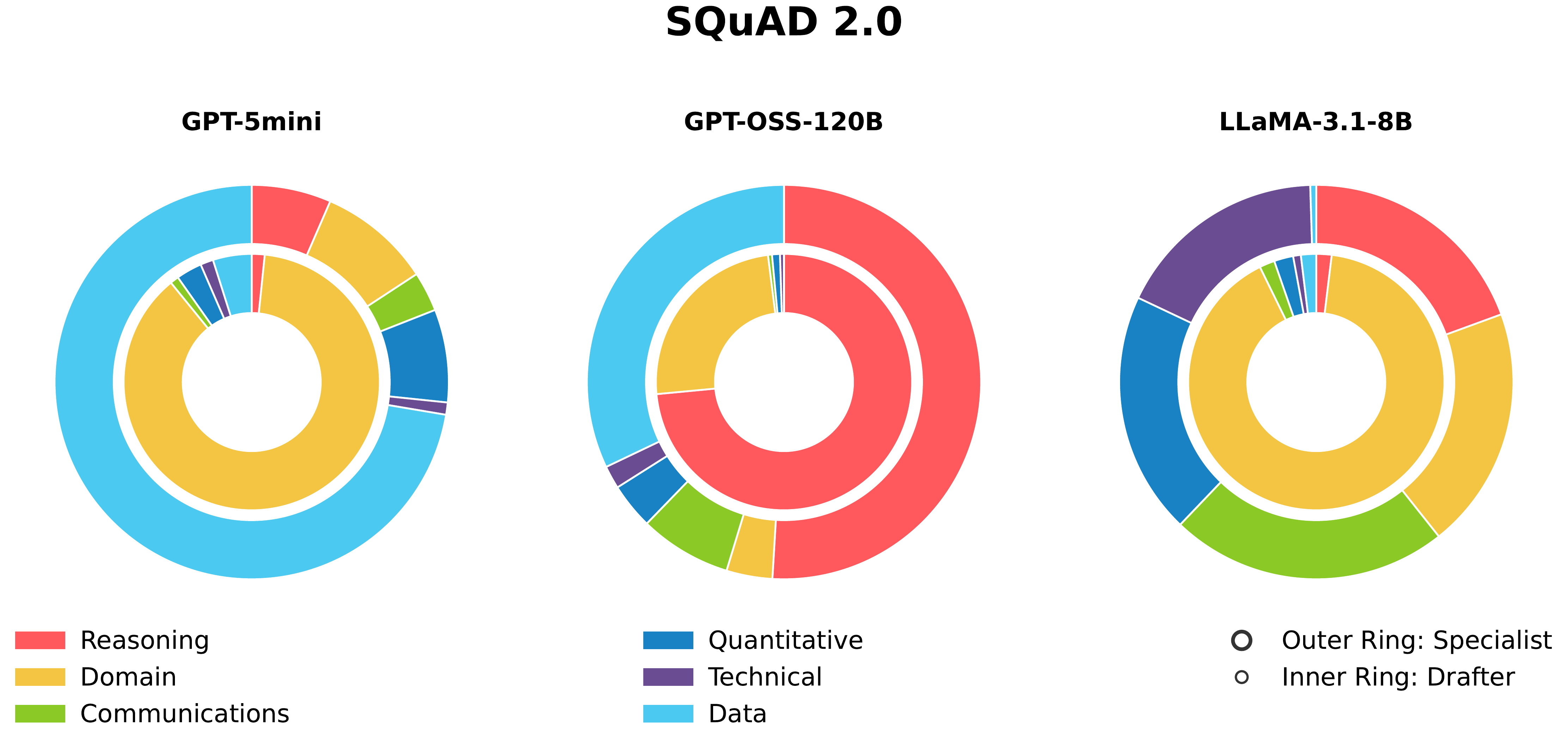}
    \caption{Skill distribution on SQuAD 2.0 across GPT-5mini, GPT-OSS-120B, and LLaMA-3.1-8B. The pie charts show the proportion of selected skill profiles under the hierarchical framework. }
    \label{fig:squad_role_distribution}
\end{figure*}

We next examine token consumption across execution policies. ~\autoref{tab:avg_token_exec_only} shows a clear and consistent pattern: STRICT is the most token-efficient policy across all benchmarks and all three models.
On MuSiQue, STRICT reduces average token usage to 3,795, 3,633, and 3,282 for GPT-5 mini, GPT-OSS-120B, and Llama-3.1-8B, respectively, far below other execution policies. 
The same pattern holds on MuSR, where STRICT again uses the fewest tokens for all three models. 
On SQuAD 2.0, the token advantage of STRICT is even more pronounced, with only 1,554, 1,539, and 1,148 average tokens, respectively. 
By contrast, NOCAP is usually the most expensive execution policy, especially on MuSiQue and MuSR, where its token usage grows substantially. 
AUTO and BALANCE generally occupy the middle range, offering moderate cost compared with STRICT and NOCAP.

These results reveal a clear performance-cost trade-off across execution policies. A detailed description of the relationships among the execution policies, particularly their differences in coordination strictness and budget constraints, is provided in Appendix~\ref{sec:Trade off}.
In contrast, STRICT consistently provides the strongest efficiency advantage, often preserving competitive performance while using only a small fraction of the token budget. Overall, these results show that execution policy serves as an effective control knob within the hierarchical framework: stricter policies favor efficiency, whereas more flexible policies can improve performance when additional coordination cost is acceptable.

\subsection{MAS Coordination Behavior}

To better understand how different organizational settings shape collective reasoning, we further analyze coordination behavior through the lens of skill-selection distributions and answer behavior.
We primarily report results on SQuAD 2.0 in the main text, and provide further analysis on MuSiQue and MuSR in Appendix~\ref{appendix:coordination_patterns}. 

As shown in~\autoref{fig:squad_role_distribution}, hierarchical coordination produces clear but model-specific specialization patterns on SQuAD 2.0.
GPT-5mini and LLaMA-3.1-8B overwhelmingly assign the drafter to the domain specialist, with selection rates of 87.50\% and 90.82\%, respectively. In contrast, GPT-OSS-120B more often assigns the drafter to the reasoning specialist, reaching 73.50\%. Specialist selection also differs substantially across backbones. GPT-5mini relies most heavily on the data specialist, which accounts for 72.28\% of specialist assignments. GPT-OSS-120B concentrates mainly on reasoning and data specialists, at 50.94\% and 32.08\%. By contrast, LLaMA-3.1-8B distributes specialist assignments much more broadly across reasoning, domain, communications, quantitative, and technical skills, indicating weaker specialization and less stable coordination. These results suggest that hierarchy provides a structured coordination mechanism, but the resulting division of labor remains strongly shaped by the underlying model. 

In~\autoref{tab:squad_unanswerable_answer_rate}, the abstention results on the unanswerable questions of SQuAD 2.0 further show how hierarchy changes system behavior in meaningful ways.
Specifically, the single-agent baseline never abstains, and flat MAS shows little to no abstention, ranging from 0 to 3.02\%.
In contrast, hierarchical execution policies raise abstention rates substantially, reaching 19.39\% to 39.78\%.
Among these execution policies, NOCAP yields the highest abstention rates for all three models, reaching 31.18\% for GPT-5mini, 39.78\% for GPT-OSS-120B, and 27.96\% for LLaMA-3.1-8B. This pattern suggests that hierarchical coordination is especially useful when the task benefits from controlled information flow, stable role assignment, and layered checking, particularly in cases where the correct behavior is to withhold an answer rather than guess.

\begin{table}[!t]
\centering
\scalebox{0.8}{
\begin{tabular}{lccc}
\hline
\multicolumn{4}{c}{\begin{tabular}[c]{@{}c@{}}\textbf{SQuAD 2.0} Unanswerable Questions\\ Abstention Rate (\%)\end{tabular}}          \\ \hline
\multicolumn{1}{l|}{} &
  \multicolumn{1}{c|}{\begin{tabular}[c]{@{}c@{}}GPT-\\ 5mini\end{tabular}} &
  \multicolumn{1}{c|}{\begin{tabular}[c]{@{}c@{}}GPT-\\ OSS-120B\end{tabular}} &
  \begin{tabular}[c]{@{}c@{}}LLaMA-\\ 3.1-8B\end{tabular} \\ \hline
\multicolumn{1}{l|}{Baseline} & \multicolumn{1}{c|}{0}              & \multicolumn{1}{c|}{0}              & 0              \\
\multicolumn{1}{l|}{Flat}     & \multicolumn{1}{c|}{3.02}           & \multicolumn{1}{c|}{0}              & 0              \\ \hline
\multicolumn{1}{l|}{AUTO}     & \multicolumn{1}{c|}{22.58}          & \multicolumn{1}{c|}{36.56}          & 19.39          \\
\multicolumn{1}{l|}{STRICT}   & \multicolumn{1}{c|}{13.98}          & \multicolumn{1}{c|}{32.26}          & 19.35          \\
\multicolumn{1}{l|}{BALANCE}  & \multicolumn{1}{c|}{30.11}          & \multicolumn{1}{c|}{35.48}          & 21.51          \\
\multicolumn{1}{l|}{NOCAP}    & \multicolumn{1}{c|}{\textbf{31.18}} & \multicolumn{1}{c|}{\textbf{39.78}} & \textbf{27.96} \\ \hline
\end{tabular}
}
\caption{AbsRate(\%) on Unanswerable Questions in SQuAD 2.0 under different organizational settings and execution policies.}
\label{tab:squad_unanswerable_answer_rate}
\end{table}

\section{Conclusion}

We presented \textbf{OrgAgent}, a company-style hierarchical MAS framework that organizes collaboration through explicit layers of governance, execution, and compliance. Across three reasoning benchmarks, we show that structuring agents like a company can often achieve a better balance between task effectiveness and token efficiency than flat coordination and single-agent baselines, with especially clear gains on MuSiQue and SQuAD 2.0. 
We further find that hierarchy also changes coordination behavior itself, often leading to more structured role allocation and verification patterns. 
Our findings suggest that company-style hierarchy provides a useful paradigm for building more capable, economical, and interpretable multi-agent systems.

\section*{Limitations}
First, although \textbf{OrgAgent} performs well on open-ended reasoning tasks, its improvements are more limited on multiple-choice benchmarks such as MMLU~\cite{hendrycks2020measuring} and MMLU-Pro~\cite{yue2025mmmu}. A possible reason is that these tasks provide a constrained answer space, leaving less room for hierarchical coordination to contribute. Second, our framework uses a fixed maximum number of discussion rounds. When agents fail to converge before this limit, the interaction is forcibly terminated, which may leave the coordination process incomplete. In such cases, hallucinated or weakly supported claims introduced by one agent may not be fully corrected, and can instead be propagated or reinforced through subsequent interaction. Although our hierarchical design includes review and compliance steps, it cannot fully eliminate this risk. Finally, we evaluate only a limited set of models, tasks, and organizational settings, and do not examine other practical factors such as latency, stability across repeated runs, or human evaluation.

\bibliography{ref}

\appendix
\newpage
\clearpage 

\section{Appendix}
\label{sec:appendix}

\subsection{Additional Details of the Benchmarks}
\label{sec:benchmark_details}

\begin{table*}[h]
\centering
\small
\setlength{\tabcolsep}{6pt}
\renewcommand{\arraystretch}{1.05}
\begin{tabular}{l|c|c}
\hline
\textbf{Benchmark} & \textbf{Type} & \textbf{Total} \\
\hline
MuSR     & Multistep soft reasoning & 756 \\
MuSiQue  & Compositional multi-hop QA & 24,814 \\
SQuAD 2.0 & Reading comprehension with unanswerable questions & 151,054 \\
\hline
\end{tabular}
\caption{Overview of the benchmarks.}
\label{tab:benchmark_overview}
\end{table*}

To complement the brief benchmark description in the main text, we provide additional details on the characteristics and scale of the three datasets used in our experiments. These benchmarks were selected because they stress different aspects of multi-agent coordination, ranging from long-context narrative reasoning to compositional evidence aggregation and answerability detection. Table~\ref{tab:benchmark_overview} summarizes the three benchmarks used in our experiments, including their task types and overall dataset sizes.

\paragraph{MuSR.}
MuSR is a benchmark for \emph{multistep soft reasoning} over long free-text narratives. Rather than focusing on short factual lookups, it requires models to combine multiple pieces of implicitly distributed evidence and perform commonsense-driven reasoning over story-like contexts. The benchmark contains three domains: \emph{murder mysteries}, \emph{object placements}, and \emph{team allocation}. According to the official benchmark description, these domains contain 250, 256, and 250 instances respectively, for a total of 756 examples. Because MuSR emphasizes long-form narrative understanding and non-trivial intermediate inference, it is especially suitable for analyzing whether hierarchical coordination helps agents organize evidence and reduce reasoning errors in complex textual settings. 

\paragraph{MuSiQue.}
MuSiQue is a compositional multi-hop question answering benchmark designed to make shortcut-based reasoning difficult. Its construction explicitly combines single-hop questions into connected multi-hop questions, so the final answer depends on evidence drawn across multiple supporting paragraphs rather than on isolated lexical overlap. The official paper reports statistics for \emph{MuSiQue-Ans}, the answerable version of the dataset: 19,938 training instances, 2,417 development instances, and 2,459 test instances, for a total of 24,814 examples. The same paper further notes that \emph{MuSiQue-Full} contains twice as many questions in each split by pairing each answerable example with an unanswerable counterpart. In our setting, MuSiQue provides a useful testbed for studying whether organizational structure improves multi-step evidence composition and answer synthesis under moderate context complexity. 

\paragraph{SQuAD 2.0.}
SQuAD 2.0 is a reading comprehension benchmark that combines standard extractive QA with adversarially written unanswerable questions. In contrast to purely answerable QA tasks, models must both extract a correct text span when one is supported by the passage and abstain when no answer is entailed. The official dataset statistics report 130,319 training examples, 11,873 development examples, and 8,862 test examples. The benchmark extends SQuAD 1.1 by adding over 50,000 unanswerable questions written to resemble answerable ones, making superficial span matching insufficient. This benchmark is particularly useful in our study because it tests whether structured coordination helps agents distinguish between answer generation and answer refusal, especially when plausible distractors are present in the context. 
\subsection{Detailed Benchmark Metric Definitions}
\label{app:benchmark_metrics}

\paragraph{Notation.}
Let $N$ denote the total number of evaluation examples, and let $i \in \{1,\dots,N\}$ index an example. For each example $i$, $\hat{y}_i$ denotes the predicted answer and $y_i$ denotes the corresponding gold answer. The indicator function $\mathbb{I}(\cdot)$ equals $1$ if its condition is true and $0$ otherwise.

\paragraph{MuSR.}
For \textbf{MuSR}, we report \textbf{Accuracy}, defined as
\begin{equation}
\mathrm{Accuracy}=\frac{1}{N}\sum_{i=1}^{N}\mathbb{I}(\hat{y}_i=y_i).
\end{equation}
This metric measures the proportion of examples for which the predicted answer exactly matches the gold answer.

\paragraph{MuSiQue and SQuAD 2.0.}
For both \textbf{MuSiQue} and \textbf{SQuAD 2.0}, we report the standard token-level \textbf{F1-score}. Let $P_i$ and $G_i$ denote the predicted and gold answer token sets for example $i$. We first compute precision and recall:
\begin{equation}
\mathrm{Precision}_i=\frac{|P_i \cap G_i|}{|P_i|},
\qquad
\mathrm{Recall}_i=\frac{|P_i \cap G_i|}{|G_i|},
\end{equation}
where $|P_i \cap G_i|$ is the number of overlapping tokens between the prediction and the gold answer, $|P_i|$ is the number of predicted tokens, and $|G_i|$ is the number of gold tokens. The example-level F1-score is
\begin{equation}
\mathrm{F1}_i=
\frac{2\cdot \mathrm{Precision}_i \cdot \mathrm{Recall}_i}
{\mathrm{Precision}_i+\mathrm{Recall}_i},
\end{equation}
and the final benchmark-level F1-score is
\begin{equation}
\mathrm{F1}=\frac{1}{N}\sum_{i=1}^{N}\mathrm{F1}_i.
\end{equation}

\paragraph{Across-run statistics.}
For each setting, we run the system $K$ times. Let $s_k$ denote the benchmark score obtained in run $k$, where $k \in \{1,\dots,K\}$. We report the mean score and standard deviation:
\begin{equation}
\bar{s}=\frac{1}{K}\sum_{k=1}^{K}s_k,
\end{equation}
\begin{equation}
\mathrm{std}(s)=
\sqrt{\frac{1}{K-1}\sum_{k=1}^{K}(s_k-\bar{s})^2}.
\end{equation}

\paragraph{Abstention rate on unanswerable questions.}
For the unanswerable subset of \textbf{SQuAD 2.0}, let $\mathcal{U}$ denote the set of unanswerable questions, let $\hat{a}_i$ denote the system output for example $i$, and let $\mathcal{N}$ denote the set of normalized no-answer outputs. The abstention rate is defined as
\begin{equation}
\mathrm{AbsRate}_{\mathrm{unans}}(\%)=
\frac{1}{|\mathcal{U}|}
\sum_{i\in\mathcal{U}}
\mathbb{I}(\hat{a}_i\in\mathcal{N})\times 100.
\end{equation}
This metric measures how often the system abstains from answering questions that do not have a valid answer in the context.

\subsection{Framework Configurations and Maximum Rounds}
Table~\ref{tab:structure_rounds} summarizes the organizational settings used in our experiments. We consider two coordination structures: a flat organization and a hierarchical organization. The flat setting does not impose explicit layered governance, and all agents interact within a single-level coordination process with a maximum of three rounds. By contrast, the hierarchical setting decomposes collaboration into distinct layers with different responsibilities. In Layer A, management agents (CEO, CTO, and COO) are responsible for high-level planning, role assignment, and execution control, with up to three rounds of governance-level coordination. In Layer B, execution is carried out under three alternative modes with different coordination depths: \textsc{DIRECT}, which produces an answer in a single round; \textsc{LIGHT MAS}, which allows lightweight iterative collaboration for up to three rounds; and \textsc{FULL MAS}, which supports deeper multi-agent interaction for up to five rounds. This design enables us to systematically vary both organizational structure and coordination depth, and to analyze how these choices affect task performance and token efficiency.

\begin{table}[t]
\centering
\small
\setlength{\tabcolsep}{4pt}
\renewcommand{\arraystretch}{1.05}
\begin{tabular}{l|c|c|c}
\hline
\textbf{Structure} & \textbf{Layer} & \textbf{Configuration} & \textbf{Max Round} \\
\hline
Flat & --  & -- & 3 \\
\hline
\multirow{4}{*}{Hierarchical}
& Layer A & CEO / CTO / COO & 3 \\
\cline{2-4}
& \multirow{3}{*}{Layer B} & DIRECT & 1 \\
\cline{3-4}
& & LIGHT MAS & 3 \\
\cline{3-4}
& & FULL MAS & 5 \\
\hline
\end{tabular}
\caption{Organizational structures and maximum coordination rounds used in our framework.}
\label{tab:structure_rounds}
\end{table}

\subsection{Relationships Between Execution Policies and Token Consumption}
\label{sec:Trade off}

\begin{figure*}[t]
    \centering
    \includegraphics[width=\textwidth]{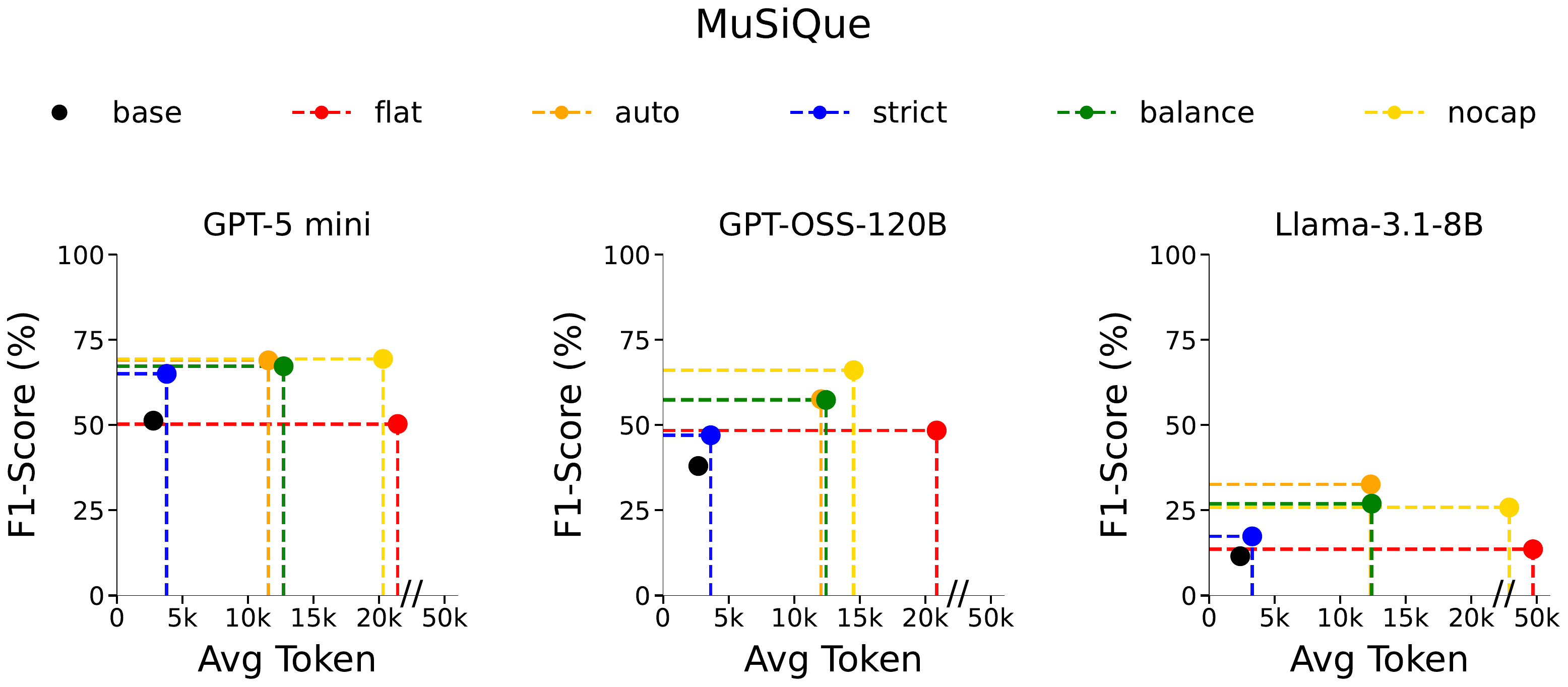}
    \caption{Token-performance trade-off on MuSiQue across GPT-5 mini, GPT-OSS-120B, and Llama-3.1-8B.}
    \label{fig:musique_tradeoff}
\end{figure*}

\begin{figure*}[t]
    \centering
    \includegraphics[width=\textwidth]{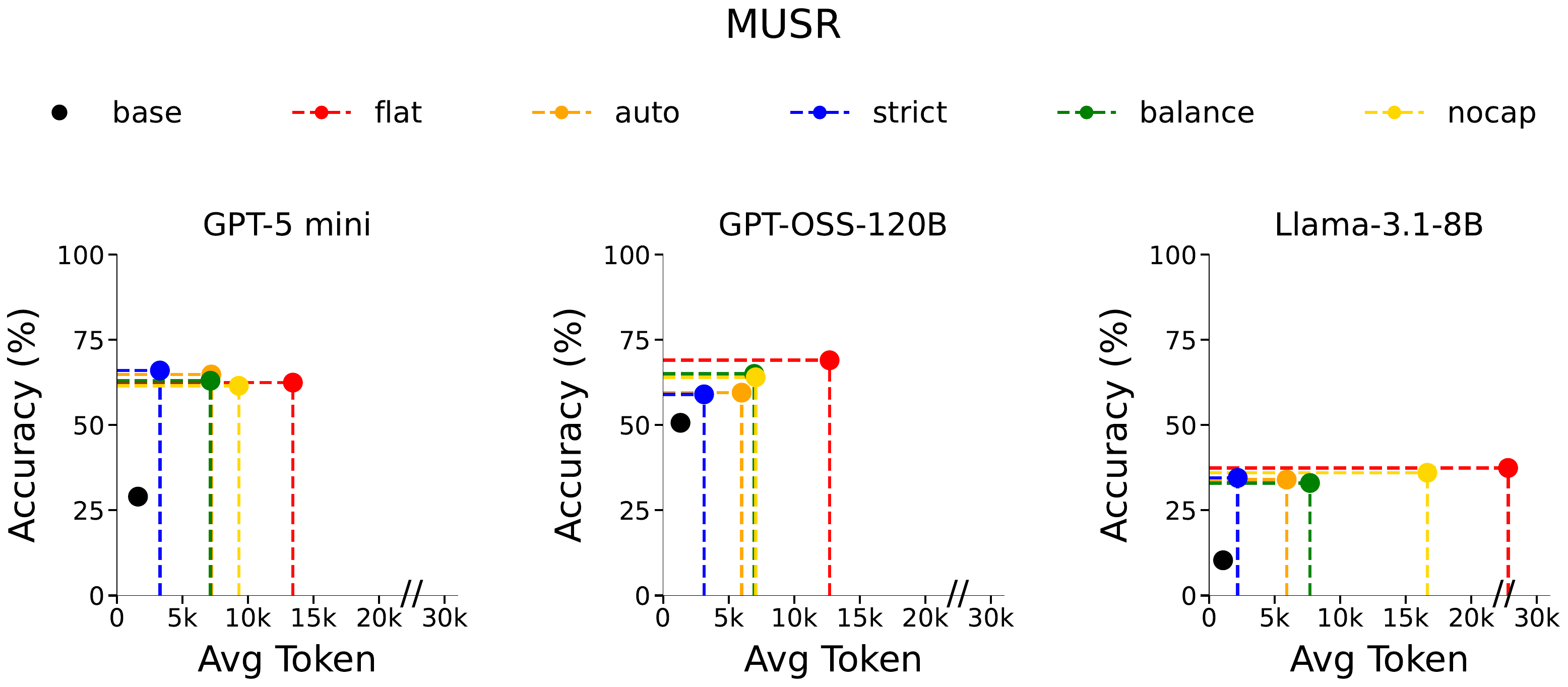}
    \caption{Token-performance trade-off on MuSR across GPT-5 mini, GPT-OSS-120B, and Llama-3.1-8B.}
    \label{fig:musr_tradeoff}
\end{figure*}

\begin{figure*}[t]
    \centering
    \includegraphics[width=\textwidth]{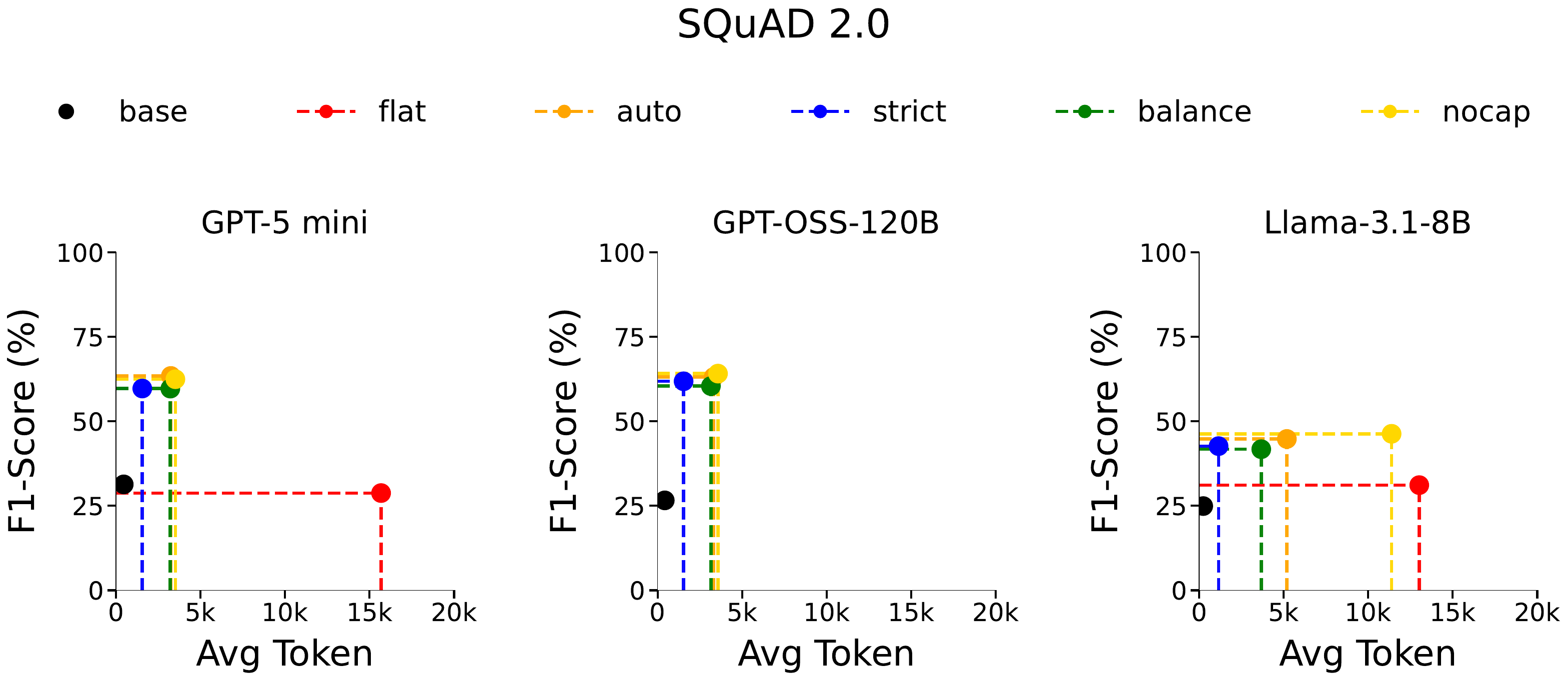}
    \caption{Token-performance trade-off on SQuAD 2.0 across GPT-5 mini, GPT-OSS-120B, and Llama-3.1-8B.}
    \label{fig:squad2_tradeoff}
\end{figure*}

This appendix provides a supplementary view of how execution policies relate to token consumption and performance. As shown in ~\autoref{fig:musique_tradeoff}, \autoref{fig:musr_tradeoff}, and \autoref{fig:squad2_tradeoff}, STRICT is consistently the most token-efficient execution policy, while NOCAP usually uses the most tokens. AUTO and BALANCE typically lie between these two extremes.

The figures also show that the performance gain from additional tokens is benchmark dependent. On MuSiQue, more flexible policies often achieve stronger results, while on MuSR and SQuAD 2.0, the performance differences among execution policies are smaller than their token differences. Overall, these results suggest that the execution policies form a spectrum from efficiency-oriented coordination to more flexible but more expensive coordination.

\section{Additional Coordination Pattern Analysis}
\label{appendix:coordination_patterns}

\begin{figure*}[t]
    \centering
    \includegraphics[width=\textwidth]{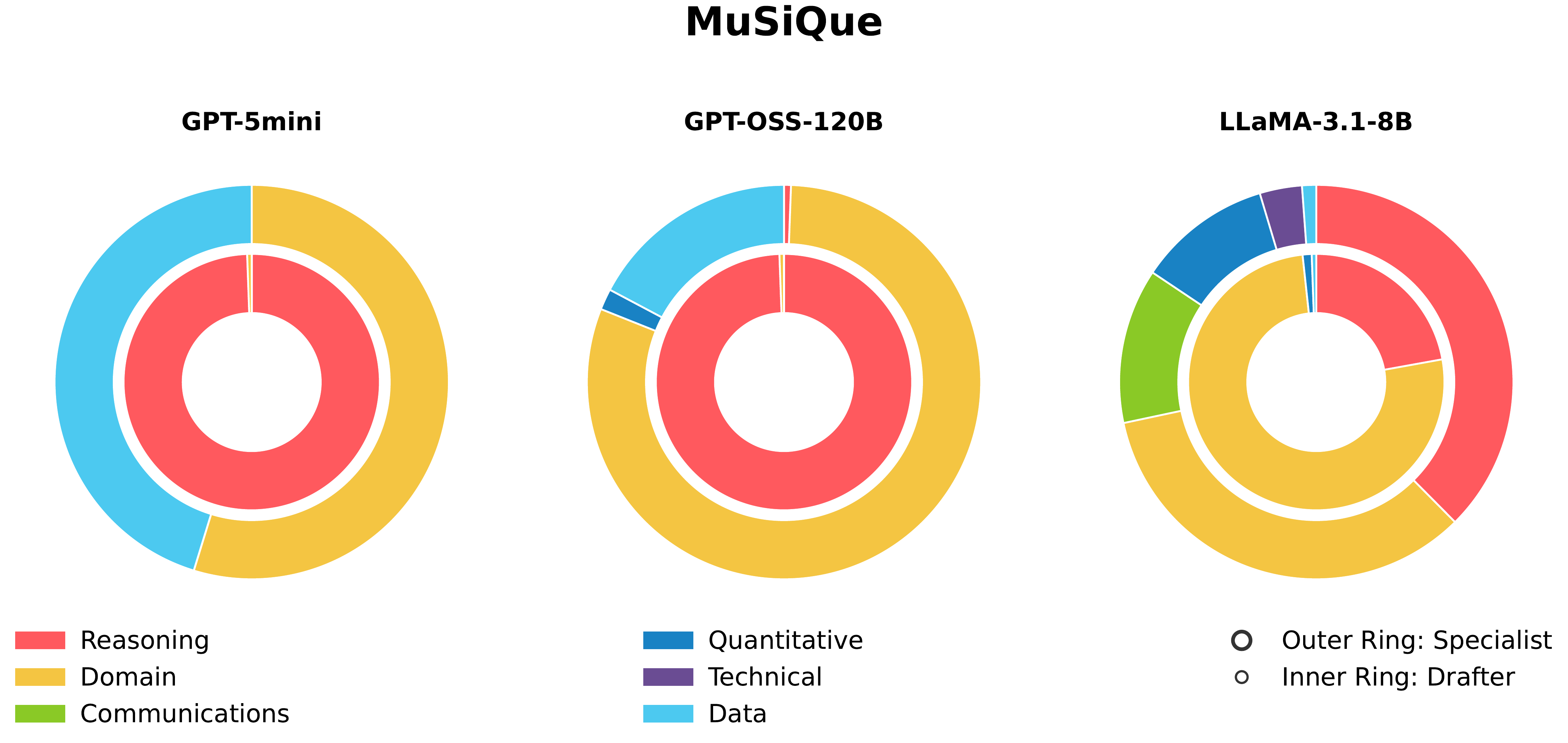}
    \caption{Skill-selection distributions across six skill types on MuSiQue. The top row shows the Drafter, and the bottom row shows the Specialist, across GPT-5mini, GPT-OSS-120B, and LLaMA-3.1-8B.}
    \label{fig:musique_pie}
\end{figure*}

\paragraph{MuSiQue.}
As shown in Figure~\ref{fig:musique_pie}, skill selection on MuSiQue also exhibits clear model-dependent patterns. For GPT-5mini and GPT-OSS-120B, the drafter is dominated by the reasoning specialist, while specialist selection is concentrated on a small subset of skills, especially domain knowledge and, in some cases, quantitative or data support. In contrast, LLaMA-3.1-8B shows a more mixed allocation for both drafter and specialist skills, with responsibility distributed across several skill types. This again suggests that hierarchical coordination induces specialization, but the sharpness and stability of this specialization depend strongly on the backbone model.

\begin{figure*}[t]
    \centering
    \includegraphics[width=\textwidth]{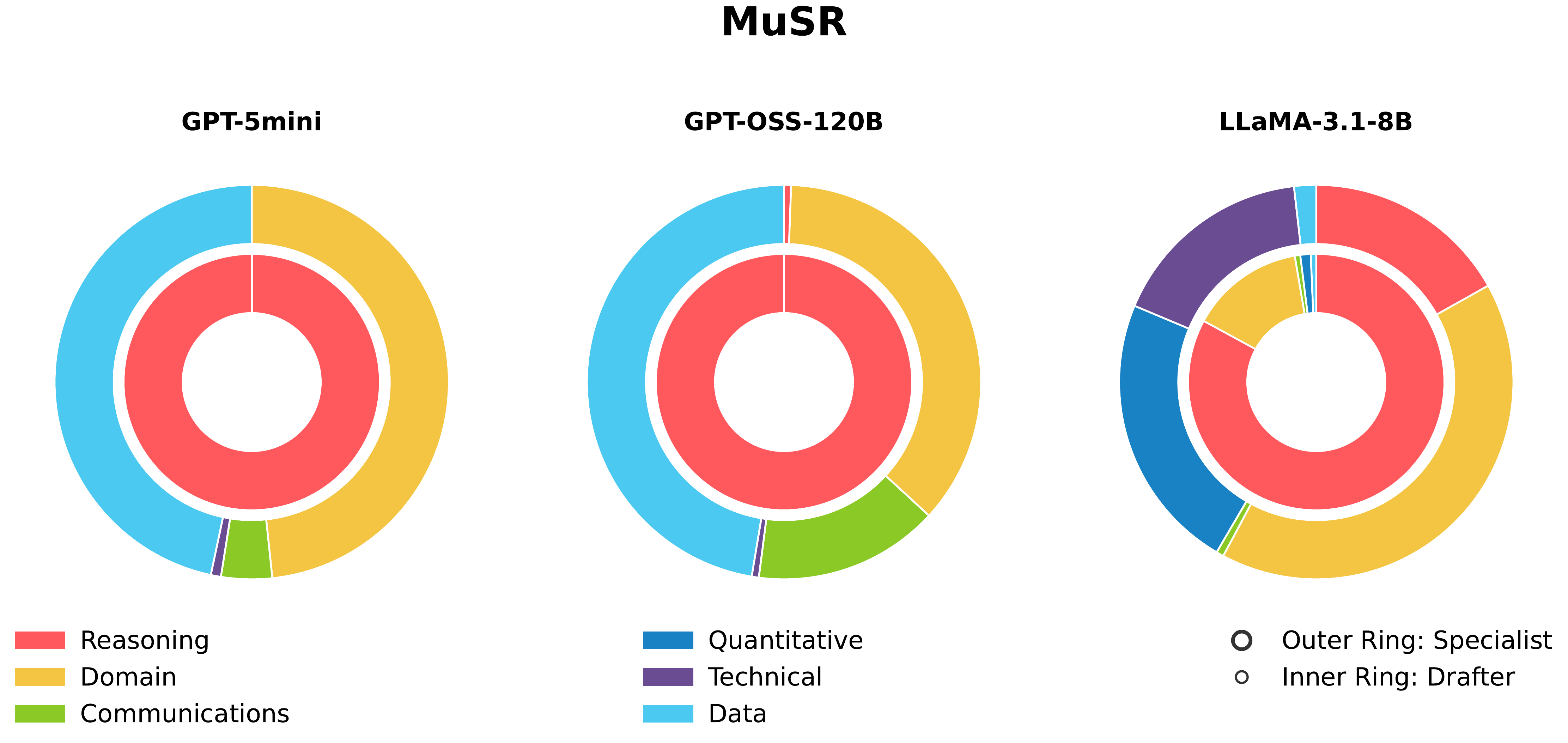}
    \caption{Skill-selection distributions across six skill types on MuSR. The top row shows the Drafter, and the bottom row shows the Specialist, across GPT-5mini, GPT-OSS-120B, and LLaMA-3.1-8B.}
    \label{fig:murs_pie}
\end{figure*}

\paragraph{MuSR.}
Figure~\ref{fig:murs_pie} shows a similar trend on MuSR. GPT-5mini and GPT-OSS-120B continue to assign the drafter primarily to reasoning-oriented agents, while specialist usage is concentrated mainly on domain and data-related support. LLaMA-3.1-8B remains comparatively more diffuse, with specialist assignments spread across multiple skills rather than concentrated on a single dominant type. Overall, the MuSR results are consistent with the main text: hierarchy provides a structured mechanism for division of labor, but the resulting coordination pattern remains strongly model-specific.

\end{document}